\documentclass[aps,reprint,groupedaddress]{revtex4-2}

\usepackage{amsmath}
\usepackage{graphicx}
\usepackage{siunitx}
\usepackage[linesnumbered,ruled]{algorithm2e}
\usepackage{url}

\begin{document}

\title{Convolutional hybrid-PIC modeling of non-neutral plasmas}

\author{Jin-Long Jiao}
\email[]{jiao.jl@zju.edu.cn}
\affiliation{Zhejiang Institute of Modern Physics, Institute of Astronomy, School of Physics, Zhejiang University, Hangzhou 310027, China}

\date{\today}

\begin{abstract}
Non-neutral plasmas can excite many nonlinear plasma phenomena, e.g., collisionless shocks, sheath layers, solitons, and plasma plumes. It is a fundamental issue in fields such as astrophysics, space physics, nuclear fusion, and plasma propulsion. Hybrid PIC methods are currently the most commonly used techniques for simulating non-neutral plasmas. However, the numerical methods for solving hybrid PIC model of non-neutral plasma have some shortcomings. Here, a new method called convolutional hybrid-PIC (conv-HPIC) is proposed to address this problem. This method replaces the Poisson's equation in the hybrid PIC with a convolution equation. The conv-HPIC method avoids iteration, suppresses numerical noise and exhibits high computational efficiency. It is expected to become a powerful tool for exploring nonlinear phenomena in non-neutral plasmas.
\end{abstract}

\maketitle

\section{Introduction}
Maintaining charge neutrality is one of the fundamental properties of plasmas, it is typically described using the quasi-neutral condition. However, when the plasma density changes drastically on the scale of the Debye length, the quasi-neutral condition breaks down, resulting in a non-neutral plasma. Typical phenomena in non-neutral plasmas include, but are not limited to, collisionless shocks, sheath layers, solitons, and plasma plumes. These phenomena are widely observed in fields such as astrophysics, space physics, nuclear fusion, and plasma propulsion. Therefore, non-neutral plasmas represent a fundamental issue with extensive implications.

This paper focus on numerical simulation methods for the non-neutral plasmas. There are three types of simulation methods for non-neutral plasmas: fully kinetic methods, fluid methods, and hybrid methods. Fully kinetic methods, such as the Particle-In-Cell (PIC) method and the Fokker-Planck method, are first-principle approaches that can self-consistently simulate the evolution of non-neutral plasmas, but they require high computational costs. Fluid methods (e.g., two-fluid method) require less computational resources, but ignore kinetic effects. Hybrid methods combine the advantages of both approaches before, retaining some kinetic effects while reducing computational costs. As a result, hybrid PIC methods are currently the most commonly used techniques for simulating non-neutral plasmas \cite{Cichocki}.

The HPIC discussed here refers specifically to a particle-ion-fluid-electron model. In this model, ions are treated kinetically, electrons are assumed as a massless fluid, and the electromagnetic fields are described using the Darwin model, which omits the displacement current term in the Ampere's law \cite{Winske}. In non-neutral plasmas, the quasi-neutral condition is no longer applicable. There are currently two methods for addressing this problem. The first method is replacing the quasi-neutral condition with the Poisson's equation. The Poisson's equation is typically solved using iterative methods, such as the Newton-Raphson or Gauss-Seidel method \cite{Lipatov,Vu,Araki}. However, these iterative methods require artificial iterations (or tolerance) and initial values, which can lead to potential program interruptions or divergence. The second method is retaining the displacement current term in the Ampere's law, which means that the Darwin model is not used. This approach ensures that the Poisson's equation is automatically satisfied within a charge conservation system \cite{Cai}. Although this method avoids the need for iteration, it requires very small time steps to ensure computational stability of the electromagnetic fields, significantly increasing computational costs. A new approach to overcoming the limitations of the aforementioned methods is expected to impact various fields of plasma physics.

Here, I propose a novel computional method, called Convolutional Hybrid PIC, for simulating non-neutral plasmas. The key to the new method is to introduce a convolution equation associated with electron density. Using the convolution equation instead of the Poisson's equation can avoid iterations and does not need to solve the displacement current term. Furthermore, this method can effectively suppress the inherent numerical noise in the HPIC method. 

\section{The Convolutional Hybrid-PIC Method}
This section is divided into three parts to introduce the Convolutional Hybrid PIC method (conv-HPIC). The first part introduces the governing equations of hybrid PIC model for the non-neutral plasma; the second part discusses how to derive the convolution equation; and the third part describes how to develop the conv-HPIC method based on the convolution equation.

\subsection{The Governing Equations of Non-neutral Plasmas for Hybrid-PIC Model}
There are two types of hybrid PIC model for the non-neutral plasmas. The first type of model uses the Darwin model rather than the Maxwell's equations to describe the electromagnetic fields, the displacement currents term $\partial\textbf{E}/\partial t$ is neglected. This model requires the Poisson's equation when describing non-neutral plasmas. The second type of model uses the Maxwell's equations to describe the electromagnetic fields, the displacement currents term is retained. One can prove that the Poisson's equation is satisfied rigorously in this type of model \cite{Cai}. The conv-HPIC model discussed here is developed based on the first method mentioned above. Below is an introduction to the first method.

The electromagnetic fields are described using the Darwin model in the Hybrid PIC method. The displacement current term is ignored in the Ampere's law. Therefore, the Maxwell's equations are simplified to the Darwin model (in c.g.s.\ unit),
\begin{eqnarray}
\frac{\partial\textbf{B}}{\partial t}=-c\nabla\times\textbf{E},\\
\nabla\times\textbf{B}=\frac{4\pi}{c}\textbf{J}=\frac{4\pi e}{c}\left(Zn_i\textbf{u}_i-n_e\textbf{u}_e\right),\\
\nabla\cdot\textbf{E}=4\pi e\left(Zn_i-n_e\right),\\
\nabla\cdot\textbf{B}=0.
\end{eqnarray}
Where $\textbf{E}$ and $\textbf{B}$ represent electric and magnetic fields. $n_i$ and $n_e$ are ion and electron densities. $\textbf{u}_i$ and $\textbf{u}_e$ are ion and electron fluid velocities. $Z$ is the ionization degree. $c$ is the speed of light. $e$ is the elementary charge. One can prove that if Eq.\ (4) holds at the initialization of the simulation, it will automatically be satisfied thereafter. Therefore, Eq.\ (4) does not need to be solved at each time step and is usually used to verify the correctness of the simulation.

The ions are regarded as pseudoparticles. The motion of ions are described by the Lorentz's equation,
\begin{eqnarray}
m_i\frac{d\textbf{v}_i}{dt}=q_i\left(\textbf{E}+\frac{1}{c}\textbf{v}_i\times\textbf{B}\right),\\
\frac{d\textbf{r}_i}{dt}=\textbf{v}_i,
\end{eqnarray}
where $\textbf{v}_i$ and $\textbf{r}_i$ are the velocity and position of the $i$-th pseudoparticle, respectively.

The electrons are assumed as a fluid. The momentum equation of electron fluid is,
\begin{equation}
m_e\frac{d\textbf{u}_e}{dt}=-e\left(\textbf{E}+\frac{1}{c}\textbf{u}_e\times\textbf{B}\right)-\frac{1}{n_e}\nabla\cdot\overline{\overline{\textbf{P}}}_e.
\end{equation}
$\overline{\overline{\textbf{P}}}_e$ is the electron pressure tensor. Ignoring the inertia of electron fluid (i.e., $m_e d\textbf{u}_e/dt\approx 0$), the momentum equation (7) can be simplified to the generalized Ohm's law,
\begin{equation}
\textbf{E}=-\frac{1}{c}\textbf{u}_e\times\textbf{B}-\frac{1}{en_e}\nabla p_e.
\end{equation}
Where assuming that the plasma is weakly collisional and the electrons are isotropic, i.e., $\overline{\overline{\textbf{P}}}_e=p_e\overline{\overline{\textbf{I}}}$. Substituting Eq.\ (2) into Eq.\ (8) and eliminating $\textbf{u}_e$, the generalized Ohm's law (8) can be rewritten as,
\begin{equation}
\textbf{E}=\frac{-1}{en_e}\nabla p_e-\frac{Zn_i}{cn_e}\textbf{u}_i\times\textbf{B}-\frac{1}{4\pi e n_e}\textbf{B}\times\left(\nabla\times\textbf{B}\right).
\end{equation}
$\textbf{u}_i$ and $n_i$ are obtained by mapping the velocity $\textbf{v}_i$ and position $\textbf{r}_i$ of the ion pseudoparticles to grids.

The Eqs.\ (1,3,5,6,9) are the governing equations of non-neutral plasmas for hybrid-PIC model. For quasi-neutral plasmas, the Poisson's equation (3) can be replaced by the quasi-neutral condition (i.e., $n_e\cong Zn_i$). The other equations will also be simplified by using the quasi-neutral condition.

In the governing equations, the electron density is obtained by solving the Poisson's equation (3) and the generalized Ohm's law (9) simultaneously. The most widely used method for solving these two equations is the iterative method, such as Newton-Raphson or Gauss-Seidel methods \cite{Lipatov,Vu,Araki}. However, iterative methods are easy to cause program interruption or divergence due to the artificial iterations (or tolerance) and initial values. Therefore, a non-iterative numerical method of solving electron density is crucial for hybrid PIC. Below introduce a interesting convolution equation for solving the electron density. This equation not only avoids iteration but also offers additional advantages. The convolution equation is the core of the conv-HPIC method.

\subsection{Derivation of Convolution Equation Associated with Electron Density}

The electron pressure $p_e$ is a function of electron density $n_e$ and temperature $T_e$ in general. So, the pressure gradient can be expanded as $\nabla p_e\left(n_e,T_e\right)=\left.\frac{\partial p_e}{\partial n_e}\right|_{T_e}\nabla n_e+\left.\frac{\partial p_e}{\partial T_e}\right|_{n_e}\nabla T_e$. Here defines $\alpha=\left.\frac{\partial p_e}{\partial n_e}\right|_{T_e}$, $\beta=\left.\frac{\partial p_e}{\partial T_e}\right|_{n_e}$ and $\varepsilon=\frac{1}{4\pi e^2n_e}$. The Eq.\ (9) can be written as,
\begin{equation}
\textbf{E}=-4\pi e\varepsilon\left(\alpha\nabla n_e+\beta\nabla T_e\right)-\textbf{F},
\end{equation}
where $\textbf{F}=\frac{4\pi e^2\varepsilon}{c}\textbf{u}_i\times\textbf{B}+e\varepsilon\textbf{B}\times\left(\nabla\times\textbf{B}\right)$.

\begin{figure}
\includegraphics[width=7cm]{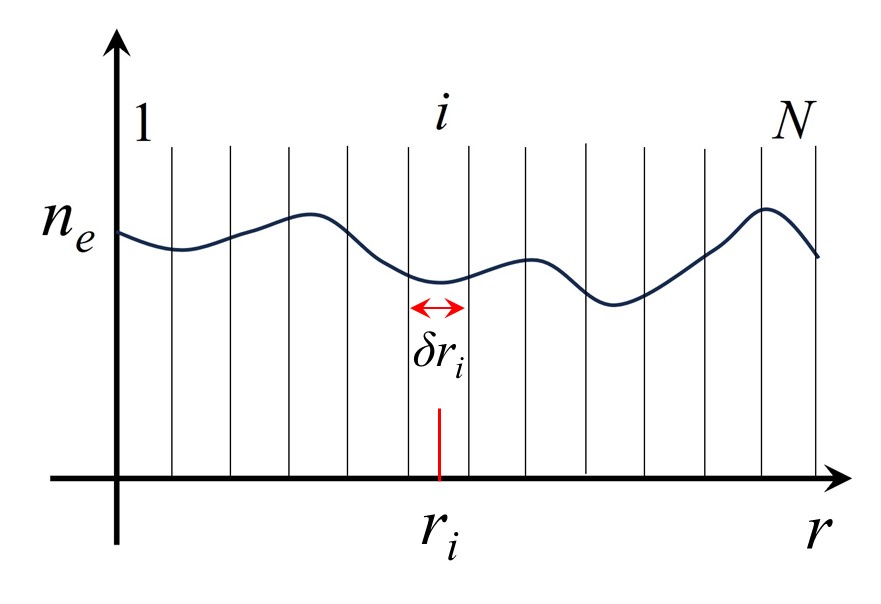}
\caption{Schematic diagram of dividing plasma (for one-dimensional case).}
\end{figure}

Divide the region where the plasma is located into several subregions $\Omega_N$, as shown in Fig.\ 1. For any subregion $\Omega_i$, its range is $\textbf{r}\in\left[\textbf{r}_i-\delta\textbf{r}_i/2,\textbf{r}_i+\delta\textbf{r}_i/2\right]$. The Eq.\ (10) on the subregion $\Omega_i$ can be written as,
\begin{eqnarray}
\Pi\left(\frac{\textbf{r}-\textbf{r}_i}{\delta\textbf{r}_i}\right)\textbf{E}&=&-\Pi\left(\frac{\textbf{r}-\textbf{r}_i}{\delta\textbf{r}_i}\right)\nonumber \\
&&\times\left[4\pi e\varepsilon\left(\alpha\nabla n_e+\beta\nabla T_e\right)+\textbf{F}\right],
\end{eqnarray}
where $\Pi\left(\frac{\textbf{r}-\textbf{r}_i}{\delta\textbf{r}_i}\right)$ is the rectangular function, and its value is 1 in the subregion $\Omega_i$, elsewhere is 0. If the subregion $\Omega_i$ is small enough (i.e.\ $\delta\textbf{r}_i\rightarrow0$), one can assume that,
\begin{equation}
\alpha\left(\textbf{r}\right)\approx\alpha\left(\textbf{r}_i\right), \beta\left(\textbf{r}\right)\approx\beta\left(\textbf{r}_i\right), \varepsilon\left(\textbf{r}\right)\approx\varepsilon\left(\textbf{r}_i\right).
\end{equation}
Thus, the Fourier transform of the first term on the right side of the Eq.\ (11) is,
\begin{eqnarray}
&&\int_{\textbf{R}}{\Pi\left(\frac{\textbf{r}-\textbf{r}_i}{\delta\textbf{r}_i}\right)\varepsilon\left(\textbf{r}\right)\alpha\left(\textbf{r}\right)\nabla n_e\left(\textbf{r}\right)e^{-\mathrm{i}\textbf{k}\cdot\textbf{r}}d\textbf{r}}\approx\nonumber \\
&&\varepsilon\left(\textbf{r}_i\right)\alpha\left(\textbf{r}_i\right)\int_{\textbf{R}}{\Pi\left(\frac{\textbf{r}-\textbf{r}_i}{\delta\textbf{r}_i}\right)\nabla n_e\left(\textbf{r}\right)e^{-\mathrm{i}\textbf{k}\cdot\textbf{r}}d\textbf{r}}=\nonumber \\
&&\varepsilon\left(\textbf{r}_i\right)\alpha\left(\textbf{r}_i\right)\Theta\ast\left[\mathrm{i}\textbf{k}n_e\left(\textbf{k}\right)\right].
\end{eqnarray}
Where $\Theta$ represents the Fourier transform of $\Pi\left(\frac{\textbf{r}-\textbf{r}_i}{\delta\textbf{r}_i}\right)$. The symbol [$\ast$] represents the convolution operator. The remaining terms in Eq.\ (11) can be transformed in the similar procedure. Finally, the Fourier transform of the Eq.\ (11) is,
\begin{eqnarray}
\textbf{E}\left(\textbf{k}\right)&=&-4\pi e\mathrm{i}\textbf{k}\left[\varepsilon\left(\textbf{r}_i\right)\alpha\left(\textbf{r}_i\right)n_e\left(\textbf{k}\right)+\varepsilon\left(\textbf{r}_i\right)\beta\left(\textbf{r}_i\right)T_e\left(\textbf{k}\right)\right]\nonumber \\
&&-\textbf{F}\left(\textbf{k}\right).
\end{eqnarray}
Where the convolutional operation $\Theta\ast$ on both sides of the Eq.\ (14) is canceled out. Similarly, the Fourier transform of the Poisson's equation (3) is, ($\Theta\ast$ is also canceled out)
\begin{equation}
\mathrm{i}\textbf{k}\cdot\textbf{E}\left(\textbf{k}\right)=4\pi e\left[n_i\left(\textbf{k}\right)-n_e\left(\textbf{k}\right)\right].
\end{equation}
The electric field $\textbf{E}\left(\textbf{k}\right)$ can be eliminated by substituting the Eq.\ (14) into (15). Then one obtains,
\begin{eqnarray}
n_e\left(\textbf{k}\right)&=&\left[n_i\left(\textbf{k}\right)-\varepsilon\left(\textbf{r}_i\right)\beta\left(\textbf{r}_i\right)k^2T_e\left(\textbf{k}\right)+\frac{1}{4\pi e}\rm{i}\textbf{k}\cdot\textbf{F}\left(\textbf{k}\right)\right]\nonumber \\
&&\times f\left(k\right).
\end{eqnarray}
Where $f\left(k\right)=\frac{1}{1+{\varepsilon\left(\textbf{r}_i\right)\alpha\left(\textbf{r}_i\right)k}^2}$, $k=\left|\textbf{k}\right|=\sqrt{k_x^2+k_y^2+k_z^2}$. Applying the inverse Fourier transform for the Eq.\ (16), the electron density in $\textbf{r}$-space can be written as a convolution form,
\begin{eqnarray}
&&n_e\left(\textbf{r},\textbf{r}_i\right)=\nonumber \\
&&\left[n_i\left(\textbf{r}\right)+\varepsilon\left(\textbf{r}_i\right)\beta\left(\textbf{r}_i\right)\nabla^2T_e\left(\textbf{r}\right)+\frac{1}{4\pi e}\nabla\cdot\textbf{F}\left(\textbf{r}\right)\right]\ast f\left(r\right)\nonumber \\
&&=\int_{\textbf{R}}\left[n_i\left(\textbf{l}\right)+\varepsilon\left(\textbf{r}_i\right)\beta\left(\textbf{r}_i\right)\nabla^2T_e\left(\textbf{l}\right)+\frac{1}{4\pi e}\nabla\cdot\textbf{F}\left(\textbf{l}\right)\right]\nonumber \\
&&\times f\left(\left|\textbf{r}-\textbf{l}\right|\right)d\textbf{l}.
\end{eqnarray}
Where $r=\left|\textbf{r}\right|=\sqrt{x^2+y^2+z^2}$. The Fourier transform of $f\left(k\right)$ can be converted to the $n$-dimensional Hankel transform due to $k^2$ is symmetrical \cite{Bracewell},
\begin{equation}
f\left(r\right)=\frac{\left(2\pi\right)^{-\frac{n}{2}}}{r^{\frac{n}{2}-1}}\int_{0}^{\infty}{f\left(k\right)J_{\frac{n}{2}-1}\left(kr\right)k^{n/2}dk}.
\end{equation}
Where $J$ is the Bessel function of the first kind. Here defines $a^2=\frac{1}{\varepsilon\left(\textbf{r}_i\right)\alpha\left(\textbf{r}_i\right)}$.

(1) For 1D case ($n$=1, $k=|kx|$, $r=|x|$),
\begin{equation}
f\left(r\right)=\frac{a}{2}e^{-ar}.
\end{equation}
One can prove that $\int_{-\infty}^{\infty}f\left(r\right)dr=1$, so $f\left(r\right)$ is normalized.

(2) For 2D case ($n$=2, $k=\sqrt{k_x^2+k_y^2}$, $r=\sqrt{x^2+y^2}$),
\begin{equation}
f\left(r\right)=\frac{a^2}{2\pi}K_0\left(ar\right).
\end{equation}
Where $K_0$ is the modified Bessel functions of second kind of orders zero. One can prove that $\int_{0}^{\infty}f\left(r\right)2\pi rdr=1$, so $f\left(r\right)$ is normalized.

(3) For 3D case ($n$=3, $k=\sqrt{k_x^2+k_y^2+k_z^2}$, $r=\sqrt{x^2+y^2+z^2}$),
\begin{equation}
f\left(r\right)=\frac{a^2}{4\pi r}e^{-ar}.
\end{equation}
One can prove that $\int_{0}^{\infty}{f\left(r\right)4\pi r^2dr}=1$, so $f\left(r\right)$ is normalized. Note that the kernel function $f\left(r\right)$ has a same form with the Debye screening potential in plasma. That means the Eq.\ (17) contains the charge separation effect self-consistently.

The Eq.\ (17) holds only in the subregion $\Omega_i$ because the assumptions (12) is used. The electron density in the whole region can be obtained by integrating all the subregions,
\begin{eqnarray}
&&n_e\left(\textbf{r}\right)=\sum_{i}\left[\Pi\left(\frac{\textbf{r}-\textbf{r}_i}{\delta\textbf{r}_i}\right)n_e\left(\textbf{r},\textbf{r}_i\right)\right]\approx \nonumber \\
&&\int_{\textbf{R}}{\delta\left(\textbf{r}-\textbf{r}_i\right)n_e\left(\textbf{r},\textbf{r}_i\right)d\textbf{r}_i}=\nonumber \\
&&\int_{\textbf{R}}\left[n_i\left(\textbf{l}\right)+\varepsilon\left(\textbf{r}\right)\beta\left(\textbf{r}\right)\nabla^2T_e\left(\textbf{l}\right)+\frac{1}{4\pi e}\nabla\cdot\textbf{F}\left(\textbf{l}\right)\right]\nonumber \\
&&\times f\left(\left|\textbf{r}-\textbf{l}\right|\right)d\textbf{l}.
\end{eqnarray}
Where the coefficient $a^2$ in the kernel function $f\left(r\right)$ is $a^2=\frac{1}{\varepsilon\left(\textbf{r}\right)\alpha\left(\textbf{r}\right)}$. Here we use a relationship $\lim\limits_{\delta\textbf{r}_i\rightarrow0}{\frac{1}{\delta\textbf{r}_i}\Pi\left(\frac{\textbf{r}-\textbf{r}_i}{\delta\textbf{r}_i}\right)}=\delta\left(\textbf{r}-\textbf{r}_i\right)$.

The Eq.\ (22) can be simplified by using the equation of state (EOS) with adiabatic approximation, i.e., $p_en_e^{-\gamma}=const.$, where $\gamma$ is the specific heat ratio. This EOS is widely used in practical simulations. Thus, one have coefficient $\beta=0$. The Eq.\ (22) can be simplified as,
\begin{equation}
n_e\left(\textbf{r}\right)=\int_{\textbf{R}}\left[n_i\left(\textbf{l}\right)+\frac{1}{4\pi e}\nabla\cdot\textbf{F}\left(\textbf{l}\right)\right] f\left(\left|\textbf{r}-\textbf{l}\right|\right)d\textbf{l}.
\end{equation}
For an electrostatic problem ($\textbf{B}=0$), the convolution equation (23) can be further simplified as,
\begin{equation}
n_e\left(\textbf{r}\right)=\int_{\textbf{R}}n_i\left(\textbf{l}\right)f\left(\left|\textbf{r}-\textbf{l}\right|\right)d\textbf{l}.
\end{equation}

Numerically solving the Eq.\ (22) requires two approximations:

(1) The electron density approximation: When using the Eq.\ (22) to calculate electron density, there are four functions $\varepsilon\left(\textbf{r}\right)$, $\beta\left(\textbf{r}\right)$, $\textbf{F}\left(\textbf{r}\right)$ and $f\left(r\right)$ should be calculated first. But all four functions are associated with the electron density $n_e$. Here are two methods to solve this problem. (i) Firstly, use the electron density at time $t$ ($n_e^{t}$) to calculate the four functions, and then use the four functions to calculate the electron density at time $t+\Delta t$ ($n_e^{t+\Delta t}$). (ii) Firstly, compute $n_e^{t+\Delta t}$ by the scheme (i), and then recalculate the four functions using the $n_e^{t+\Delta t}$. Finally, use the updated four functions to recalculate the $n_e^{t+\Delta t}$. Scheme (ii) offers higher accuracy but requires more computational resources.

(2) The kernel function approximation: The kernel function's domain spans entire $\textbf{r}$-space, implying that the convolution should be performed across the entire $\textbf{r}$-space. However, one can observe that the kernel function's value exponentially decays, allowing for acceptance of a finite domain. I recommend a domain size $r=4/a$, where the value of the kernel function has decreased by two orders of magnitude ($\sim e^{-4}$).

\subsection{The Convolutional Hybrid-PIC Method}

The Eqs.\ (1,5,6,9,22) are the governing equations of the conv-HPIC method. Comparing with the governing equations disscussed in section A, the Poisson's equation (3) is replaced by the convolution equation (22).

\begin{widetext}

\begin{algorithm}[H]
\DontPrintSemicolon
\caption{Solution Procedure of Convolutional Hybrid-PIC Method}
  \textbf{Initialization}\;
  \While{$t<t_{end}$}{
          \quad Update simulation time: $t=(N+1)\Delta t$ \;
      \textbf{1.Update particles}\;
          \tcc{Using the leap-frog algorithm.}
          \quad Compute ion velocity: $\textbf{v}_i^{N+1/2}=\textbf{v}_i^{N-1/2}+\frac{Ze\Delta t}{m_i}\left(\textbf{E}^N+\frac{1}{c}\textbf{v}_i^N\times\textbf{B}^N\right)$ \;
          \quad Compute ion position: $\textbf{r}_i^{N+1}=\textbf{r}_i^{N}+\textbf{v}_i^{N+1/2}\Delta t$ \;

      \textbf{2.Mapping particles to grids}\;
          \tcc{$w_k$ is the weight of k-th particle; $\phi(\textbf{r})$ is the shape factor of particle.}
          \quad Compute ion density: $n_i^{N+1}=\sum\limits_{k=1}^{N_p} w_k\phi(\textbf{r}_k^{N+1})$ \;
          \quad Compute ion fluid velocity: $\textbf{u}_i^{N+1/2}=\sum\limits_{k=1}^{N_p} w_k\textbf{v}_k^{N+1/2}\phi(\textbf{r}_k^{N+1/2})/n_i^{N+1/2}$ \;

      \textbf{3.Update electron density}\;
        \tcc{Using the EOS of adiabatic ideal gas.}
        \quad Compute ion density: $n_e^{N+1}=[n_i^{N+1}+\frac{1}{4\pi e}\nabla\cdot\textbf{F}^{N}]\ast f^N$ \;

      \textbf{4.Update electron pressure}\;
        \tcc{Using the EOS of adiabatic ideal gas.}
        \quad Compute electron pressure: $p_e^{N+1/2}=p_{e0}(n_e^{N+1/2}/n_{e0})^\gamma$\;
        
      \textbf{5.Update electric field and magnetic field}\;
        \tcc{Compute electric field and magnetic field using the CAM-CL scheme.}
        \quad $\textbf{B}^{N+1/2}=\textbf{B}^N-\frac{c\Delta t}{2}\nabla\times\textbf{E}^N$ \;
        \quad $\textbf{E}^{N+1/2}=-\frac{\nabla p_e^{N+1/2}}{n_e^{N+1/2}}-\frac{\textbf{u}_i^{N+1/2}\times\textbf{B}^{N+1/2}}{c}+\frac{(\nabla\times\textbf{B}^{N+1/2})\times\textbf{B}^{N+1/2}}{4\pi n_e^{N+1/2}}=G(\textbf{B}^{N+1/2},n_e^{N+1/2},\textbf{u}_i^{N+1/2})$ \;
        \quad $\textbf{B}^{N+1}=\textbf{B}^{N+1/2}-\frac{c\Delta t}{2}\nabla\times\textbf{E}^{N+1/2}$ \;
        \quad $\textbf{E}^*=G(\textbf{B}^{N+1},n_e^{N+1},\textbf{u}_i^{N+1/2})$ \;
        \quad $\textbf{u}_i^*=\textbf{u}_i^{N+1/2}+\frac{Ze\Delta t}{2m_i}(\textbf{E}^*+\frac{1}{c}\textbf{u}^{N+1/2}\times\textbf{B}^{N+1})$ \;
        \quad $\textbf{E}^{N+1}=G(\textbf{B}^{N+1},n_e^{N+1},\textbf{u}_i^*)$ \;
  }
\end{algorithm}

\end{widetext}

Here, I proposed an algorithm for solving the conv-HPIC equations. There are 5 steps in each cycle:

(1) Update particles: Solve the ion motion equations under control of electromagnetic field, that is, solve the Eqs.\ (5,6). Compute new velocity and position of each ion. There are many methods for solving the ion motion equations, such as the leap-frog and the Boris' algorithm \cite{Birdsall}.

(2) Mapping particles to grids: Compute ion density and ion fluid velocity by mapping weight and velocity of each ion pseudoparticle onto the nodes of grids. The meaning of weight is that a pseudoparticle represents the actual number of ions in each cell. In the mapping process, it is necessary to consider the shape of the pseudoparticle cloud, which can be represented by a function about the particle position $\phi(\textbf{r}_k)$.

(3) Update electron density: Solve the convolution equation (22) to obtain the electron density. If use the EOS of adiabatic ideal gas, only the Eq.\ (23) needs to be solved. Furthermore, if consider the non-magnetization problem, only the Eq.\ (24) needs to be solved.

(4) Update electron pressure: In general, to compute the electron pressure, it is necessary to solve both the electron internal energy equation and the EOS. However, if an EOS for isothermal or adiabatic ideal gases is applicable, the electron pressure can be directly determined from the EOS without the need to solve the electron internal energy equation.

(5) Update electric field and magnetic field: Solve the Ampere’s law (1) and the generalized Ohm’s law (9) to obtain $\textbf{E}$ and $\textbf{B}$. There are many methods for solving these two equations, such as the CAM-CL and predictor-corrector methods \cite{Winske}.

\section{Test Example}

In this section, a simplified conv-HPIC case, an 1D collisionless electrostatic shock (CES), is selected to verify the effectiveness of the conv-HPIC method. The CES arises from the electrostatic ion-ion acoustic instability in a supersonic counterstream plasma system \cite{Moiseev,Forslund,Kato,Zhang}, and has been observed in various plasma experiments \cite{Taylor,Jiao2}. The width of the shock region is comparable to the Debye length, leading to the breakdown of the quasi-neutral condition within this region. Hence, the CES provides an ideal test environment for the non-neutral plasmas. Next, I performed a comparative study on the CES using different simulation methods to validate the conv-HPIC method.

The CES is generated in a colliding plasmas system. The two colliding plasmas are consist of fully ionized hydrogen and have identical parameters. The electron density is set to \SI{E20}{cm^{-3}}, while the electron and ion temperatures are \SI{40}{keV} and \SI{1}{keV}, respectively. The colliding speed is 1$\%$ of the speed of light. The Mach number is $Ma=u/c_s=u/\sqrt{ZT_e/Am_p}=1.53>1$, confirming the shock formation condition. For the simulation settings, the domain range is from 0 to \SI{100}{\mu m}, and the colliding plasmas initially make contact at $x$=\SI{50}{\mu m}. The total simulation time is \SI{5}{ps}. I apply an injector boundary for the plasmas and an open boundary for electric and magnetic fields. Note that no additional heating in the plasmas. Therefore, the plasma can be considered as adiabatic ideal gas, and its EOS is
\begin{equation}
p_en_e^{-\gamma}=const,
\end{equation}
where the specific heat ratio is $\gamma=\frac{f+2}{f}$, and $f$ is the degree of freedom. For 1D case, $f$=1, corresponds to $\gamma$=3.

In this section, three methods (quasi-neutral HPIC, conv-HPIC, and fully kinetic PIC methods) are used to simulate the CES. Below is a brief introduction on the three methods in context of electrostatic simulations.

\textbf{(a) Electrostatic quasi-neutral HPIC}

In electrostatic case, the magnetic field is ignored, so the generalized Ohm's law (9) is simplified as,
\begin{equation}
\textbf{E}=\frac{-1}{en_e}\nabla p_e.
\end{equation}
Using the EOS (25) and the Clapeyron equation of the ideal gas $p_e=n_ek_BT_e$, the gradient of the electron pressure is,
\begin{equation}
\nabla p_e=\gamma k_BT_e\nabla n_e.
\end{equation}
Substituting the Eq.\ (27) and the quasi-neutral condition $n_e\cong n_i$ into the Eq.\ (26), the electrostatic field is $\textbf{E}=-4\pi e\gamma\lambda_D^2\nabla n_i$, where $\lambda_D=\left(\frac{k_BT_e}{4\pi n_ee^2}\right)^{1/2}$ is the Debye length. Therefore, the governing equations of the quasi-neutral HPIC are,
\begin{eqnarray}
m_i\frac{d\textbf{v}_i}{dt}=q_i\textbf{E},\\
\textbf{E}=-4\pi e\gamma\lambda_D^2\nabla n_i.
\end{eqnarray}
The source code of the quasi-neutral HPIC in electrostatic case sees in github link: \url{https://github.com/jljiao/conv-HPIC/blob/main/HPIC1D/HPIC1D.m}

\textbf{(b) Electrostatic conv-HPIC}

The convolution equation is simplified as the Eq.\ (24) by using the EOS (25), i.e., $n_e=n_i\ast f\left(r\right)$. Where the coefficient $a^2$ in the kernel function $f\left(r\right)$ is $a^2=\frac{1}{\gamma\lambda_D^2}$. Therefore, the governing equations of the conv-HPIC are,
\begin{eqnarray}
m_i\frac{d\textbf{v}_i}{dt}=q_i\textbf{E},\\
\textbf{E}=-4\pi e\gamma\lambda_D^2\nabla n_e,\\
n_e=n_i\ast f\left(r\right).
\end{eqnarray}
The source code of the conv-HPIC in electrostatic case sees in github link: \url{https://github.com/jljiao/conv-HPIC/blob/main/HPIC1D/HPIC1D.m}

\textbf{(c) Electrostatic fully kinetic PIC}

The fully kinetic PIC method is a first principal method, providing highly accurate simulation results, but requiring expensive computations. Here it is used as a benchmark to validate the other two methods. In the fully kinetic PIC, both electrons and ions are treated as the pseudoparticles. The charge densities are obtained by mapping these particles to grids. The electrostatic potential is then calculated by solving the Poisson's equation \cite{Birdsall}. The governing equations of the fully kinetic HPIC are,
\begin{eqnarray}
m_e\frac{d\textbf{v}_e}{dt}=-e\textbf{E},\\
m_i\frac{d\textbf{v}_i}{dt}=q_i\textbf{E},\\
\nabla^2\phi=4\pi e\left({n_e-n}_i\right),\\
\textbf{E}=-\nabla\phi.
\end{eqnarray}
The source code of the fully kinetic PIC in electrostatic case sees in github link: \url{https://github.com/jljiao/conv-HPIC/blob/main/PIC1D/ESPIC1D.m}

\begin{figure}
\includegraphics[width=8cm]{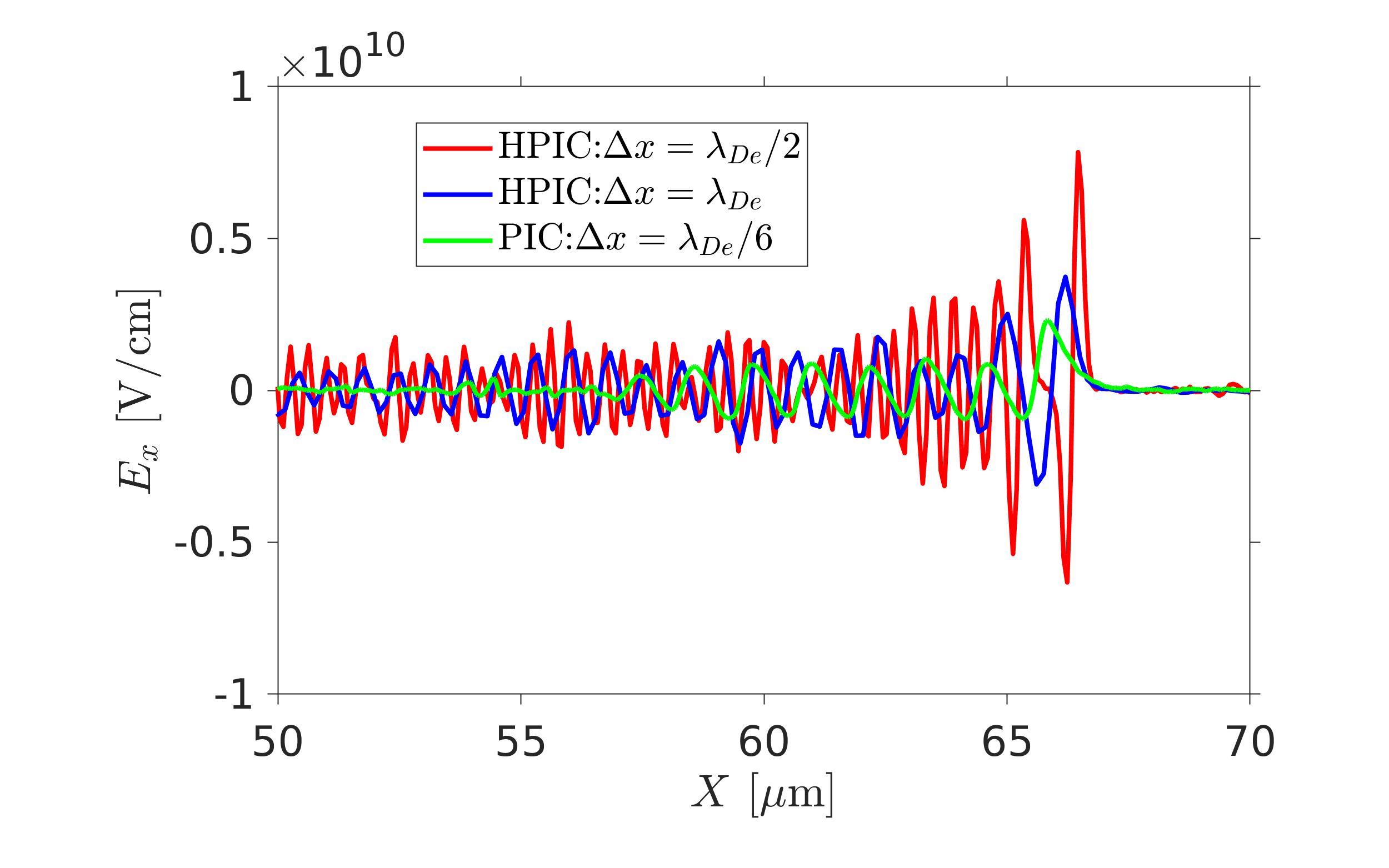}
\caption{Electric fields of collisionless electrostatic shock in quasi-neutral HPIC and fully kinetic PIC simulations at \SI{5}{ps}.}
\end{figure}

\begin{figure}
\includegraphics[width=8cm]{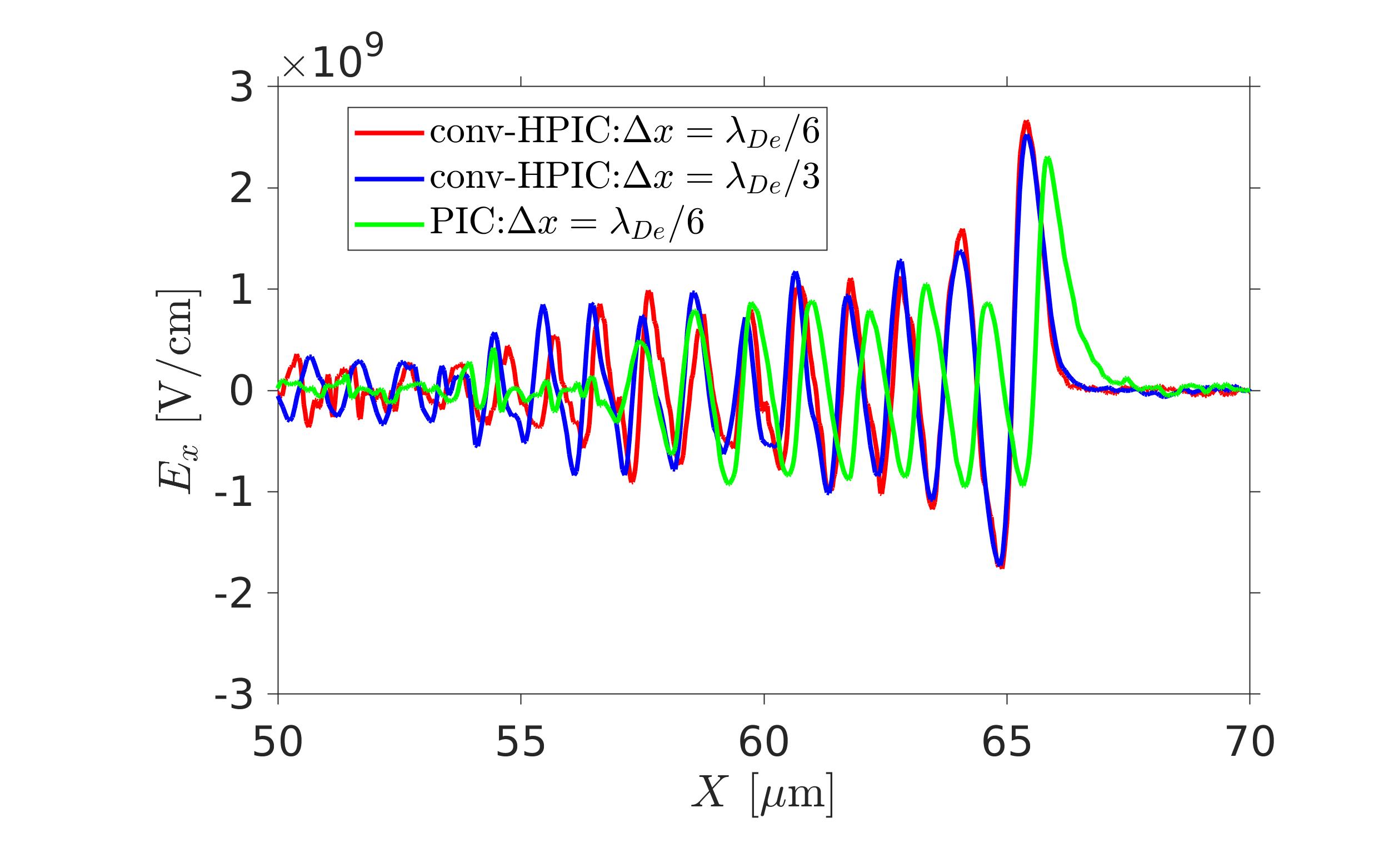}
\caption{Electric fields of collisionless electrostatic shock in conv-HPIC and fully kinetic PIC simulations at \SI{5}{ps}.}
\end{figure}

\begin{figure}
\includegraphics[width=8cm]{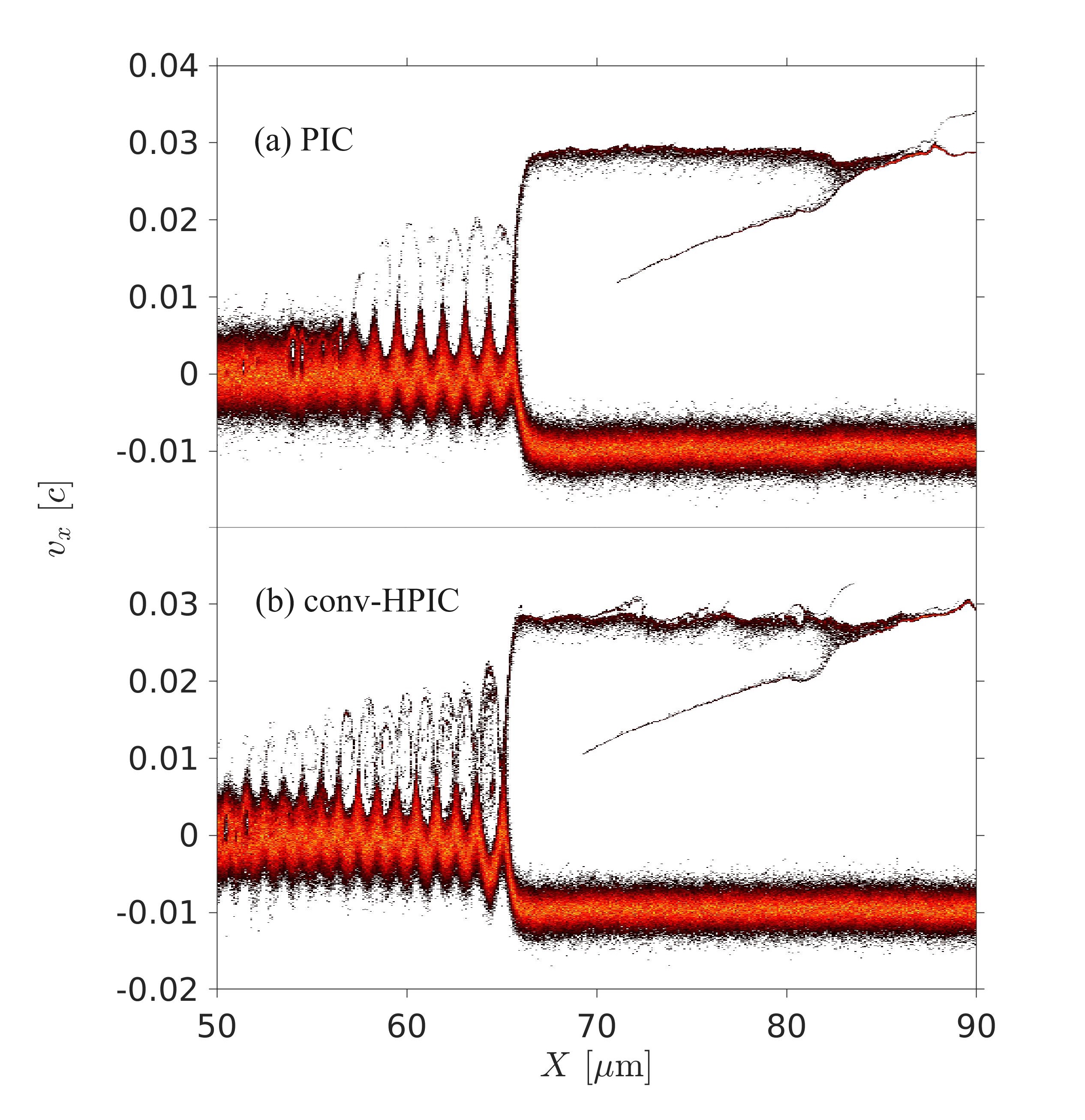}
\caption{Ion densities (arbitrary unit) in $x-v_x$ space of fully kinetic PIC (a) and conv-HPIC (b) simulations at \SI{5}{ps}.}
\end{figure}

The CES is a typical phenomenon in non-neutral plasmas. The CES is divided into upstream and downstream regions by the shock front. For example, in Fig.\ 4a, the shock front is located at $x=$\SI{66}{\mu m}, with $x<$\SI{66}{\mu m} representing the downstream region and $x>$\SI{66}{\mu m} representing the upstream region. At the shock front, the plasma density undergoes several-fold changes on the scale of the Debye length. This dramatic density variation causes significant deviations in electron and ion densities, leading to the breakdown of quasi-neutral condition and the formation of an electric field at the shock front. The electric field at the shock front can reflect a portion of the lower-energy upstream ions back toward the upstream region of the shock while slowing the remaining incoming ions, allowing them to enter the downstream region. The impact of incoming ions excites an ion acoustic wave in the downstream. The ion acoustic wave exhibits periodic oscillations in density, velocity, electric field, and ion phase space distribution.

The periodic characteristics of ion acoustic waves can be analyzed through the dispersion relation. Applying the linearization and the Fourier transform to the two-fluid equations, if retaining the Poisson's equation, the dispersion relation of ion acoustic waves associated with the charge separation effect can be obtained,
\begin{equation}
\omega^2=k^2\frac{k_B}{m_i}\left(\gamma_i T_i+\frac{\gamma_e T_e}{1+\gamma_ek^2\lambda_{D}^2}\right).
\end{equation}
On the other hand, for quasi-neutral plasmas, the charge separation effect can be ignored, so the Poisson's equation in the two-fluid equations can be replaced with the quasi-neutral condition. The resulting dispersion relation of ion acoustic waves is,
\begin{equation}
\omega^2=k^2\frac{k_B}{m_i}\left(\gamma_i T_i+\gamma_e T_e\right).
\end{equation}
The dispersion relations (37) and (38) are consistent under the long wavelength approximation $k\lambda_{D}\ll 1$ but differ under short wavelength approximation $k\lambda_{D}\gg 1$. This implies that the quasi-neutral HPIC method cannot accurately simulate the CES.

Numerical simulations show that the conv-HPIC method can accurately simulate the CES, whereas the quasi-neutral HPIC method fails to that. As the simulation grid size $\Delta x$ decreases, ion acoustic modes of short-wavelength $k=2\pi/\Delta x$ are excited. Assuming ion temperature is very low $T_i\approx 0$, under the short wavelength approximation, the Eq.\ (37) becomes $\omega=\omega_{pi}=\left(k_B T_e/m_i\lambda_D^2\right)^{1/2}$, and the Eq.\ (38) becomes $\omega=k(\gamma_ek_BT_e/m_i)$. This indicates that for short wavelength ion acoustic modes, the ion acoustic frequency approaches the ion plasma frequency $\omega_{pi}$ when the charge separation effect are considered, while there is no upper limit for the frequency if ignore this effect. Therefore, for quasi-neutral HPIC simulations, the oscillation frequency of the electric field increases as the grid size $\Delta x$ decreases, as shown in Fig.\ 2. In contrast, for conv-HPIC simulations which considering the charge separation effect, the oscillation frequency remains nearly constant regardless of grid size, as shown in Fig.\ 3. Comparing both simulation results with fully kinetic PIC, it is evident that the conv-HPIC method accurately captures the electrostatic field distribution of the CES. Moreover, comparing the ion phase space diagrams from conv-HPIC and fully kinetic PIC shows good agreement in key features, such as ion acoustic oscillations, ion Landau damping, and reflection ions, as shown in Fig.\ 4. Overall, the comparative study of quasi-neutral HPIC, conv-HPIC, and fully kinetic PIC demonstrates that the conv-HPIC method is able to accurately simulates non-neutral plasmas.

The conv-HPIC method offers several advantages over other methods when simulating non-neutral plasmas. (1) Avoiding iteration: The convolution equation (22) can be explicit solved, avoids iterative calculation. (2) High computational efficiency: Compared to the fully kinetic PIC method, the conv-HPIC method permits larger time steps \cite{Jiao1}. For the CES example, the conv-HPIC method can use a time step $\Delta t \sim \omega_{pi}^{-1}$, whereas the fully kinetic PIC method requires a much smaller time step $\Delta t \sim \Delta x/c$. In addition, convolution operation have excellent parallel performance on GPUs, which has been validated in the field of the machine learning. (3) Suppressing numerical noise: Particle methods always introduce strong numerical noise during mapping particles to grids. When numerical noise grows rapidly, it may result in non-physical simulation results. The convolution operation in the Eq.\ (22) enables the smoothing of grid quantities, significantly reducing numerical noise.

\section{Conclusion}

In summary, a new method called conv-HPIC has been proposed for simulating non-neutral plasmas. This method replaces the Poisson's equation in the HPIC with a convolution equation related to electron density. The CES simulation shows that conv-HPIC accurately captures charge separation effect in non-neutral plasmas. Advantages of this method include eliminating the need for iteration, high computational efficiency, and reduced numerical noise. It is set to become a valuable tool for exploring nonlinear phenomena in non-neutral plasmas, such as collisionless shocks, sheath layers, solitons, and plasma plumes.


\begin{acknowledgments}

This work is supported by the National Science Foundation of China (No.\ 12147103) and the Fundamental Research Funds for the Central Universities (No.\ 226-2022-00216).

\end{acknowledgments}


\begin{thebibliography}{99}
%
\bibitem{Cichocki}
F.~Cichocki, A.~Dominguez-Vazquez, M.~Merino, E.~Ahedo, Hybrid 3d model for the
  interaction of plasma thruster plumes with nearby objects, Plasma Sources
  Science and Technology 26 (2017) 125008.

\bibitem{Winske}
D.~Winske, L.~Yin, N.~Omidi, Hybrid simulation codes: Past, present and
  future—a tutorial, Space plasma simulation (2003) 136--165.

\bibitem{Lipatov}
A.~S. Lipatov, The hybrid multiscale simulation technology: an introduction
  with application to astrophysical and laboratory plasmas, Springer Science
  and Business Media, 2013.

\bibitem{Vu}
H.~X. Vu, An adiabatic fluid electron particle-in-cell code for simulating
  ion-driven parametric instabilities, Journal of Computational Physics 124(2)
  (1996) 417--430.

\bibitem{Araki}
S.~J. Araki, R.~S. Martin, D.~Bilyeu, J.~W. Koo, Sm/murf: Current capabilities
  and verification as a replacement of afrl plume simulation tool coliseum,
  52nd AIAA/SAE/ASEE Joint Propulsion Conference (2016).

\bibitem{Cai}
H.~Cai, X.~Yan, P.~Yao, Hybrid fluid–particle modeling of shock-driven
  hydrodynamic instabilities in a plasma, Matter and Radiation at Extremes 6(3)
  (2021).

\bibitem{Bracewell}
R.~Bracewell, P.~B. Kahn, The fourier transform and its applications, American
  Journal of Physics 34(8) (1966) 712.

\bibitem{Birdsall}
C.~K. Birdsall, A.~B. Langdon, Plasma physics via computer simulation, CRC
  press, 2018.

\bibitem{Moiseev}
S.~S. Moiseev, R.~Z. Sagdeev, Collisionless shock waves in a plasma in a weak
  magnetic field, Journal of Nuclear Energy 5(1) (1963) 43.

\bibitem{Forslund}
D.~W. Forslund, C.~R. Shonk, Formation and structure of electrostatic
  collisionless shocks, Physical Review Letters 25(25) (1970) 1699.

\bibitem{Kato}
T.~N. Kato, H.~Takabe, Electrostatic and electromagnetic instabilities
  associated with electrostatic shocks: Two-dimensional particle-in-cell
  simulation, Physics of Plasmas 17(3) (2010).

\bibitem{Zhang}
W.~Zhang, H.~Cai, S.~Zhu, The formation and dissipation of electrostatic shock
  waves: the role of ion–ion acoustic instabilities, Plasma Physics and
  Controlled Fusion 60(5) (2018) 055001.

\bibitem{Taylor}
R.~J. Taylor, D.~R. Baker, H.~Ikezi, Observation of collisionless electrostatic
  shocks, Physical Review Letters 24(5) (1970) 206.

\bibitem{Jiao2}
J.~L. Jiao, S.~K. He, H.~B. Zhuo, Experimental observation of ion–ion
  acoustic instability associated with collisionless shocks in laser-produced
  plasmas, The Astrophysical Journal Letters 883(2) (2019) L37.

\bibitem{Jiao1}
J.~L. Jiao, Ion current screening modeling of the ion–weibel instability, The
  Astrophysical Journal 924(2) (2022) 89.
  
\end{thebibliography}

\end{document}